\documentclass[aps,prd,twocolumn,nofootinbib,showpacs,superscriptaddress]{revtex4-1}

\usepackage{amsmath,amsfonts,amssymb,bm}
\usepackage{graphicx}
\usepackage{subfigure}
\usepackage{color}

\usepackage[colorlinks=true, pdfstartview=FitV, linkcolor=purple, citecolor= purple, urlcolor=blue]{hyperref}
\usepackage[nameinlink]{cleveref}

\definecolor{purple}{rgb}{0.5,0,0.5}
\definecolor{blue}{rgb}{0.0,0,0.9}

\newcommand{\tr}{{\text{tr}}}
\newcommand{\imag}{\text{i}}

\usepackage{xcolor}

\begin{document}

\title{Constructing the Equation of State of QCD in a functional QCD based scheme}

\author{Yi Lu}
\email{qwertylou@pku.edu.cn}
\affiliation{Department of Physics and State Key Laboratory of Nuclear Physics and Technology, Peking University, Beijing 100871, China}

\author{Fei Gao}
\email[Corresponding author: ]{fei.gao@bit.edu.cn}
\affiliation{School of Physics, Beijing Institute of Technology, Beijing 100081, China}

\author{Baochi Fu }
\email{fubaochi@pku.edu.cn}
\affiliation{Department of Physics and State
Key Laboratory of Nuclear Physics and Technology, Peking
University, Beijing 100871, China}

\author{Huichao Song}
\email{huichaosong@pku.edu.cn}
\affiliation{Department of Physics and State Key Laboratory of
Nuclear Physics and Technology, Peking University, Beijing 100871, China}
\affiliation{Center for High Energy Physics, Peking University, Beijing 100871, China}
\affiliation{Collaborative Innovation Center of Quantum Matter, Beijing 100871, China.}

\author{Yu-{X}in Liu }
\email{yxliu@pku.edu.cn}
\affiliation{Department of Physics and State Key Laboratory of
Nuclear Physics and Technology, Peking University, Beijing 100871, China}
\affiliation{Center for High Energy Physics, Peking University, Beijing 100871, China}
\affiliation{Collaborative Innovation Center of Quantum Matter, Beijing 100871, China.}

\date{\today}

\begin{abstract}
We construct the equation of state (EoS) of QCD based on the finite chemical potential information from the functional QCD approaches, with the assistance of the lattice QCD EoS.  The obtained EoS is consistent with the up-to-date estimations of the QCD phase diagram, including  a phase transition temperature at zero chemical potential of $T=155$ MeV, the curvature of the transition line  $\kappa=0.016$ and also a  critical end point at $(T,\mu_B)=(118, 600)$ MeV. In specific, the phase diagram mapping is  achieved by incorporating  the order parameters into the EoS, namely the dynamical quark mass for the chiral phase transition together with the Polyakov loop parameter for the deconfinement phase transition. We also implement the EoS in hydrodynamic simulations to compute the  particle yields, ratios and collective flow, and find that our obtained EoS agrees well with the commonly used one based on the combination of lattice QCD simulation and hadron resonance gas model.

\end{abstract}


\maketitle

\section{Introduction}

One of the main goals of the relativistic heavy ion collision experiments is to search for the critical end point (CEP) of the QCD matter and to explore the thermodynamic properties of the strong interaction matter at finite temperature and chemical potential~\cite{Akiba:2015jwa,Luo:2017faz,HADES:2019auv,Lovato:2022vgq,ALICE-USA:2022glt,Arslandok:2023utm}. To build the bridge between these measurements and the theoretical studies, the key element is the equation of state (EoS) of the QCD matter~\cite{Ding:2017giu,Guenther:2020jwe,Monnai:2021kgu,Philipsen:2021qji,Karsch:2022opd,Ratti:2022qgf,Sorensen:2023zkk}, since the EoS plays a crucial role as an input in hydrodynamic simulations for the evolution of the fireballs produced in heavy-ion collisions.

On the theoretical side, the QCD phase diagram has been widely studied via effective models~\cite{Buballa:2003qv, Fukushima:2013rx,Schaefer:2007pw,Schaefer:2008ax,Fu:2007xc,Shao:2011fk,Xin:2014ela,He:2013qq,Chelabi:2015gpc,Kojo:2020ztt,Chen:2020ath}, and the QCD approaches like  lattice QCD simulation~\cite{Borsanyi:2020fev,HotQCD:2018pds,Bonati:2018nut}, functional QCD methods including Dyson-Schwinger equations~\cite{Roberts:2000aa,Qin:2010nq,Fischer:2014ata,Fischer:2018sdj,Gunkel:2021oya,Gao:2020fbl,Gao:2020qsj} and functional Renormalization group approach~\cite{Fu:2019hdw,Dupuis:2020fhh,Fu:2022gou}. Among them, the functional QCD methods have delivered a comprehensive investigation on the phase diagram and especially, on the location of the CEP at large chemical potential. The current computations have given the phase transition line which is consistent with the lattice QCD simulation at small chemical potential and provided an estimation of  the CEP at about $\mu_{B}^{\textrm{CEP}} \approx 600$ to $650\,\textrm{MeV}$~\cite{Gunkel:2021oya,Fu:2019hdw,Gao:2020fbl,Gao:2020qsj}.
However, there is still no complete computation for the EoS which matches these up-to-date results for the  phase transition line and is accessible for the hydrodynamics simulations.  The estimated CEP is not even incorporated in the commonly applied EoS.

In this article, we then take advantage of different theoretical approaches to obtain an improved construction on 
the EoS. It mainly takes the functional QCD results about the phase transition line at finite density, which include the critical temperature, the curvature and the location of the CEP at large chemical potential, and also with the assistance of the lattice EoS at zero chemical potential. The constructed EoS is also required to  satisfy the constraints from the current results of the thermodynamic quantities. All the elements in the construction are formalized analytically, making the further application of the EoS accessible and convenient.  We also incorporate our EoS into hydrodynamic simulations to compute the experimental observables like particle yields, their ratios and  collective flow.

The article is organized as follows: In Sec.\,\ref{sec2}, we present the framework of constructing and parameterizing the EoS. In Sec.\,\ref{sec3}, we present the results of EoS in the ($\mu$, $T$) plane, and also the Maxwell construction in the first order phase transition which stablizes the evolution. Then in Sec.\,\ref{sec4}, we present the results of particle yields and ratios together with the elliptic flow $v_2$ after incorporating with the hydrodynamics simulation.  In Sec.\,\ref{sec5}, we summarize the main results and make further discussions and outlook.

\section{Construction of the EoS in a functional QCD-based scheme}\label{sec2}

Lattice QCD simulations have delivered  solid computations at vanishing chemical potential~\cite{Karsch:2001vs, Borsanyi:2010cj,Bazavov:2014pvz}. However,  due to the sign problem, it is difficult for lattice QCD to reach real chemical potentials directly. The Taylor expansion approach~\cite{Bazavov:2020bjn,Borsanyi:2018grb,Bazavov:2017dus,Guenther:2017hnx} or alternative expansion scheme~\cite{Parotto:2023xco,Bollweg:2022fqq,Bollweg:2022rps,Borsanyi:2021sxv} are also limited in small chemical potential region, and the latest lattice results can only cover up to $\mu_B/T \leq 3.5$.
On the other hand, the large chemical potential region is accessible by continuum QCD methods, i.e. functional QCD approaches including Dyson-Schwinger equations (DSEs)~\cite{Binosi:2009qm,Roberts:2000aa,Eichmann:2016yit, Fischer:2018sdj} and the functional renormalization group (fRG) method~\cite{Pawlowski:2005xe,Dupuis:2020fhh}. Therefore, one may combine the advantages of the two methods. In this work this is done by taking the following inputs: the lattice EoS data at zero chemical potential, the phase transition line data which is consistent between the lattice and also functional QCD results at small chemical potential, and also a predicted location for the critical end point from functional QCD calculations.

To calculate the EoS of the QCD matter at finite temperature $T$ and quark chemical potentials $\boldsymbol{\mu} = \{\mu_{q}\}$, we apply the integral relation between the QCD pressure $P$ and the quark number densities $\{n_{q}\}$~\cite{Chen:2012zx,Isserstedt:2020qll,Gao:2021nwz}:
\begin{equation}\label{eq:Pnq}
    P(T,\boldsymbol{\mu}) = P(T,\boldsymbol{0}) + \sum_{q} \int_{0}^{{\mu}_{q}} n_{q}(T,\mu) \, \textrm{d} \mu,
\end{equation}
where $P(T,\boldsymbol{0})$ is from the lattice QCD input. Here it is determined from the parameterized formula of the lattice QCD's trace anomaly $I(T)=(\epsilon-3P)/T^4$ at vanishing chemical potential in Ref.~\cite{Borsanyi:2010cj}:
\begin{equation}
  I(T) = e^{-h_1 /t - h2/t^2} \cdot \left[ h_0 + f_0 \cdot \frac{ \tanh(f_1 t +f_2) + 1}{1+g_1 t+g_2 t^2} \right],
\end{equation}
with $t=T/(200\,\mathrm{MeV})$, $h_0 = 0.1396$, $h_1=-0.1800$, $h_2=0.0350$, $f_0=2.76$, $f_1=6.79$, $f_2=-5.29$, $g_1=-0.47$, $g_2=1.04$ for 2+1 flavor.
The pressure at vanishing chemical potential is then obtained with:
\begin{equation}
  P(T,\boldsymbol{0})/T^{4} = \int_{0}^{T} \textrm{d}T' (I(T')/T').
\end{equation}

Therefore, the key quantity for the EoS at finite chemical potential is the quark number density.
It is related to the quark propagator $S_q$, which is the central element in the functional QCD approach:
\begin{equation}
n_{q} = - T \sum_n \int \frac{\textrm{d}^{3}\boldsymbol{p}}{(2\pi)^{3}} \tr_{C,D}[ \gamma_4 S_{q}(\omega_{n},\boldsymbol{p})]\, ,
\end{equation}
with $\omega_{n}$ the Matsubara frequencies and $\boldsymbol{p}$ the spacial momentum; the trace is taken over the color index and the Dirac structure. One can further take the vanishing momentum approximation for the inverse quark propagator as:
\begin{equation}
  S_q^{-1} \simeq \imag(\omega_{n} + \imag \mu_q +  gA_{0} )\gamma_{4} + \imag \boldsymbol{\gamma}\cdot\boldsymbol{p} + M_{q} \, ,
\end{equation}
with $M_{q}$ the dynamical quark mass and $A_{0}$ the gluon condensate~\cite{Fister:2013bh,Fischer:2013eca} which relates to the Polyakov loop $\Phi$:
\begin{equation}\label{eq:A0def}
A_{0} =  \frac{2\pi T}{g} \phi \, \tau^{3},\quad \Phi = \frac{1}{3}\left[1 + 2 \cos(\pi\varphi)\right] \, .
\end{equation}
Here for simplicity we take the gluon condensate only with the Cartan generator $\tau^{3}$ following the Ref.~\cite{Fischer:2013eca},
with the eigenvalues $\phi_{\textrm{fund}} = \{\pm\varphi/2,0\}$.
From this, a simple analytical form is available for the number density, which has great advantage on the further applications.
Such construction is the main purpose of this article.
The quark number density $n_{q}$ in the $(T,\mu_{q})$ plane can be expressed by the dynamical mass $M_{q}$ and additionally,
the Polyakov loop parameter from Eq.~(\ref{eq:A0def}), as follows~\cite{Fu:2015gl}:
\begin{gather}
n_{q}(T,\mu_{q}) = 2 N_{c} \int \frac{\mathrm{d}^{3}\boldsymbol{k}}{(2\pi)^{3}} \left[ f_{q}^{+}(\boldsymbol{k};T,\mu_{q}) - f_{q}^{-}(\boldsymbol{k};T,\mu_{q}) \right], \label{eq:nq_constit} \\
f_{q}^{\pm} = \frac{\Phi(T,\mu_{q}) x_{\pm}^{2} + 2\Phi(T,\mu_{q}) x_{\pm} + 1}{x_{\pm}^{3} + 3\Phi(T,\mu_{q}) x_{\pm}^{2} + 3\Phi(T,\mu_{q}) x_{\pm} + 1}, \\[1mm]
x_{\pm}(\boldsymbol{k};T,\mu_{q}) = \exp \left[ (E_q(\boldsymbol{k};T,\mu_{q}) \mp \mu_{q})/T \right], \\[1mm]
E_{q}(\boldsymbol{k};T,\mu_{q}) = \sqrt{\boldsymbol{k}^{2} + M_{q}^{2}(T,\mu_{q})} .
\end{gather}

The next step is to quantitatively incorporate the phase diagram into the two order parameters $M_{q}(T,\mu_{q})$ and $\Phi(T,\mu_{q})$.
First of all, it has been shown from the lattice and functional QCD studies, that the chiral phase transition line takes the parametrized form as:
\begin{equation}
  \frac{T_{c}(\mu_{B})}{T_{c}(0)} =  1 - \kappa \left( \frac{\mu_{B}}{T_{c}(0)} \right)^{2} + \lambda \left( \frac{\mu_{B}}{T_{c}(0)} \right)^{4} + \cdots \, .  \label{eq:TcmuB} 
\end{equation}
For the (2+1)-flavor case, the state-of-the-art results are $T_{c}(0) = 155\;$MeV, $\kappa = 0.016$ and $|\lambda| \lesssim 10^{-4}$~\cite{Borsanyi:2020fev,HotQCD:2018pds,Bonati:2018nut,Fu:2019hdw,Gao:2020fbl}.
In addition, functional QCD approaches estimate that the CEP locates at about $\mu_{B}^{\textrm{CEP}} \approx 600$ to $650\,\textrm{MeV}$~\cite{Fu:2023lcm,Gao:2020fbl,Gunkel:2021oya,Fu:2019hdw}. This gives then a strong constraint on the order parameter of chiral phase evolution. In the present work we simply take $\mu_{B}^{\textrm{CEP}} = 600\,\textrm{MeV}$.

\begin{figure}[t]
  \centering
  \includegraphics[width=0.43\textwidth]{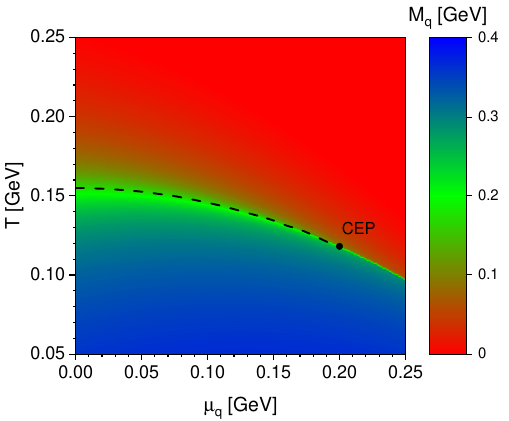}\\
  \includegraphics[width=0.43\textwidth]{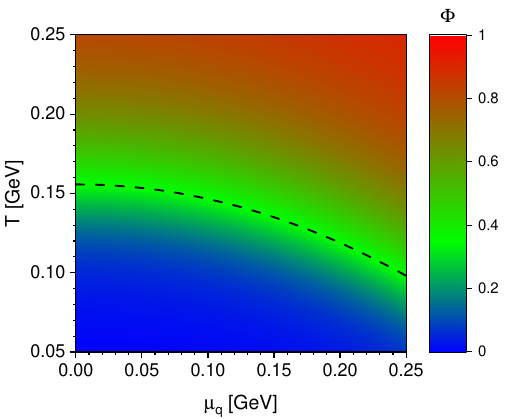}
\vspace*{-3mm}
  \caption{Density plots for the two order parameters $M_{q}$ and $\Phi$ as functions of temperature $T$ and chemical potential $\mu_q$ for light-flavor quarks.} \label{fig:orderParam}
\end{figure}

The chiral symmetry is related to the dynamical mass generation of quarks which have been continuously verified in functional QCD approaches and also effective QCD models. The quark mass at vanishing momentum can be then regarded as the order parameter of chiral phase transition~\cite{Qin:2010nq,Gao:2016qkh}.
In the vacuum, the typical mass scale for the light-flavor quarks is $M_{0} = 350\;$MeV which is suggested by lattice QCD~\cite{Bowman:2005vx,Oliveira:2016muq} and functional QCD~\cite{Fu:2019hdw,Gao:2021wun} calculations.
At finite temperature and chemical potential, one can take the Ising parametrization~\cite{Parotto:2018pwx,Dore:2022qyz} in which a CEP is included:
\begin{equation}\label{eq:MT}
  M_{q}(T,\mu_{q}) = \frac{M_{0}}{2} \left[ 1 - \mathcal{M}_{\mathrm{Ising}}(T,\mu_B) \right].
\end{equation}
%
The $(T,\mu_B)$ dependence of the order parameter can be mapped to the Ising parameter $(r,h)$, and the order parameter can be parameterized as:
\begin{align}
  \mathcal{M}_{\mathrm{Ising}} &= \mathcal{M}_{0} R^{\beta} \theta, \label{eq:IsingOP} \\
  h &= h_{0} R^{\beta\delta} \tilde{h}(\theta), \label{eq:IsingOPh} \\
  r &= R(1-\theta^{2}), \label{eq:IsingOPr}
\end{align}
with the typical parameters $\beta = 0.326$, $\delta = 4.80$, $h_0 =0.394$ and $\tilde{h}(\theta) = \theta (1-0.76201 \, \theta^2 + 0.00804 \, \theta^{4})$ as given in Ref.~\cite{Parotto:2018pwx}, and $\mathcal{M}_{0}$ is the normalization constant.
The complete mapping procedure is $(T,\mu_{B}) \rightarrow (R,\theta) \rightarrow (r,h)$.
In order to match the phase transition line Eq.~(\ref{eq:TcmuB}) precisely, we propose a non-linear map as follows:
\begin{equation}\label{eq:Isingmap}
  \begin{aligned}
  & \frac{\mu_{B} - \mu_{B}^{\mathrm{CEP}}}{\mu_B^{\mathrm{CEP}}} = - r \omega \rho \cos{\alpha_{1}} - h \omega \cos{\alpha_{2}}, \\
  & \frac{T-T^{\mathrm{CEP}}}{T^{\mathrm{CEP}}} =  f_{\mathrm{PT}}^{} (r) + h \omega \sin{\alpha_{2}},
\end{aligned}
\end{equation}
with $\alpha_{1} = \tan^{-1} ( 2\kappa \mu_{B}^{\mathrm{CEP}} /T_{c}(0) ) = 7.0^{\circ} $ and $\alpha_{2} = \alpha_{1} + 90^{\circ}$ in our case. The mapping function $f_{\mathrm{PT}}^{}$ is calibrated so that at $h=0$, Eqs.~(\ref{eq:Isingmap}) becomes exactly the parametric equations of the phase transition line in Eq.~(\ref{eq:TcmuB}), which requires
\begin{equation}\label{eq:nlmap}
  f_{\mathrm{PT}}^{}(r) = \frac{T_{c}(0)}{T^{\mathrm{CEP}}} \left[ 1 - \kappa \left(\frac{\mu^{\mathrm{CEP}}}{T_{c}(0)}\right)^{2} \left( 1 - r \omega \rho \cos {\alpha_{1}} \right)^{2} \right] - 1 .
\end{equation}
In other words, here we consider $h$ as the ``distance" towards the phase transition line, and $r$ being the projection coordinate on the phase transition line. The unconstrained parameters here are $\omega$ and $\rho$, which are currently set the same as those reported in Ref.~\cite{Parotto:2018pwx}. To sum up, the parameters for the dynamical mass are listed in Table~\ref{tab:param_m} and the respective temperature and chemical potential dependence of dynamical quark  mass $M_{q}$ is depicted in the upper panel of Fig.~\ref{fig:orderParam}.
With such a setup the temperature and chemical potential dependence of the chiral order parameter is consistent with the results from lattice QCD simulations and functional QCD methods.

\begin{table}[htb]
\centering
\caption{The mapping parameters of the dynamical mass in Eqs.~\labelcref{eq:MT,eq:IsingOP,eq:IsingOPh,eq:IsingOPr,eq:Isingmap}.
for the light-flavor quarks, including the position of the CEP.}
\begin{ruledtabular}
%
\begin{tabular}{cccc}\label{tab:param_m}
   $\omega$ & $\rho$ & $\mathcal{M}_{0}$ & $(T^{\textrm{CEP}},\mu_B^{\textrm{CEP}})\;$[MeV]   \\
  \cline{1-4}
   $1.0$ & $2.0$ & $0.75$ &  $(118.1,600)$  \\
\end{tabular}
\end{ruledtabular}
\end{table}

While $M_{q}$ is the order parameter of chiral phase transition and is parametrized above,  the Polyakov loop parameter $\Phi$ is usually connected to the deconfinement transition. Here we bypass the argument of the relation between Polyakov loop and deconfinement, and simply consider $\Phi$ as an order parameter for another phase transition in the gluon sector.

We apply the Polyakov loop data at zero $\mu_q$ from Ref.~\cite{Fu:2015gl}, which is denoted as $\Phi(T,0) = L(t)$ with $t = T/T_c(0)$. The following fit function is used here as:
\begin{equation}
  L(t) = 2 \left[ 1 + \exp \left( \frac{1+l_1 t^3}{m_1 t + m_2 t^6} \right) \right]^{-1},
\end{equation}
with $l_1 = 2.732$, $m_1 = 0.5495$ and $m_2 = 1.831$.
For the Polyakov loop at finite $\mu_{B}^{}$, we follow the expansion scheme proposed in~\cite{Borsanyi:2021sxv,Mondal:2021jxk,Mukherjee:2021tyg}, which suggests a temperature scaling behaviour of the thermodynamic functions:
\begin{gather}\label{eq:PhiT}
  \Phi(T,\mu_q) = L(t_{\Phi}), \\
  t_{\Phi} = \frac{T}{T_{c}(0)}+\kappa \left(\frac{3\, \mu_{q}}{T_{c}(0)}\right)^{2} \, .
\end{gather}
Here we choose  $T_{c}(0)$ and $\kappa$ for the deconfinement phase transition  the same  as those for the chiral phase transition as shown in Fig.~\ref{fig:orderParam}. Note that the two phase transition lines are not necessarily identical which may induce some new phases like quarkyonic phase~\cite{Philipsen:2022wjj,Philipsen:2019qqm}.
Since the QCD phase transition is mainly characterized by the chiral phase transtion, deviation of the two phase transitions may only induce some delicate changes in the observables which require further investigations in the future.

In short, we set up a functional QCD framework for the EoS by mapping the up-to-date knowledge of the QCD phase structure to the order parameters. The free parameters here are those in the map functions, namely $\omega$, $\rho$ and $\mathcal{M}_{0}$. As we will demonstrate in the following, the obtained EoS is comparable with the lattice QCD prediction and the experimental observations quantitatively, which has a weak dependence on the free parameters.

\section{Numerical results of the EoS with first order phase transition}\label{sec3}

Next, we consider the $(2+1)$-flavor case, the baryon, electric charge and strangeness chemical potentials $(\mu_{B}^{},\mu_{Q}^{},\mu_{S}^{})$ are associated with the quark chemical potentials as:
\begin{align}
  \mu_{u} &= \frac{1}{3}\mu_{B}^{} + \frac{2}{3}\mu_{Q}^{}, \label{eq:BS_muu} \\
  \mu_{d} &= \frac{1}{3}\mu_{B} - \frac{1}{3}\mu_{Q}^{}, \label{eq:BS_mud} \\
  \mu_{s} &= \frac{1}{3}\mu_{B} - \frac{1}{3}\mu_{Q}^{} - \mu_{S}^{}, \label{eq:BS_mus}
\end{align}

\begin{figure}[b]
  \centering
  \includegraphics[width=0.22\textwidth]{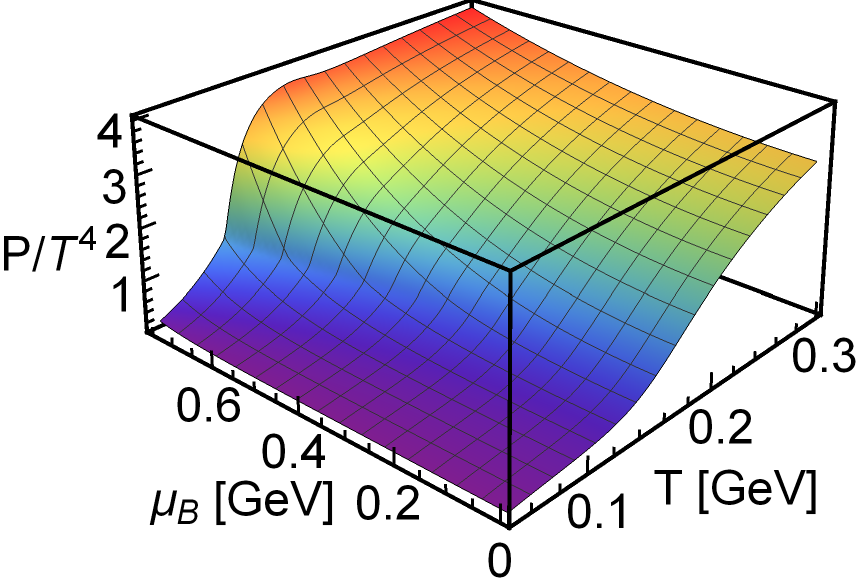} \hspace*{1mm}
  \includegraphics[width=0.22\textwidth]{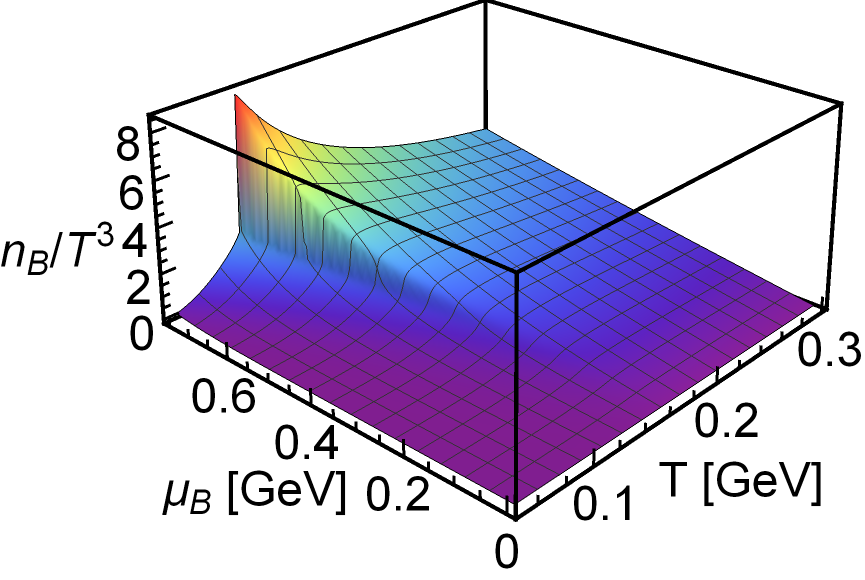}\\[1pt]
  \includegraphics[width=0.22\textwidth]{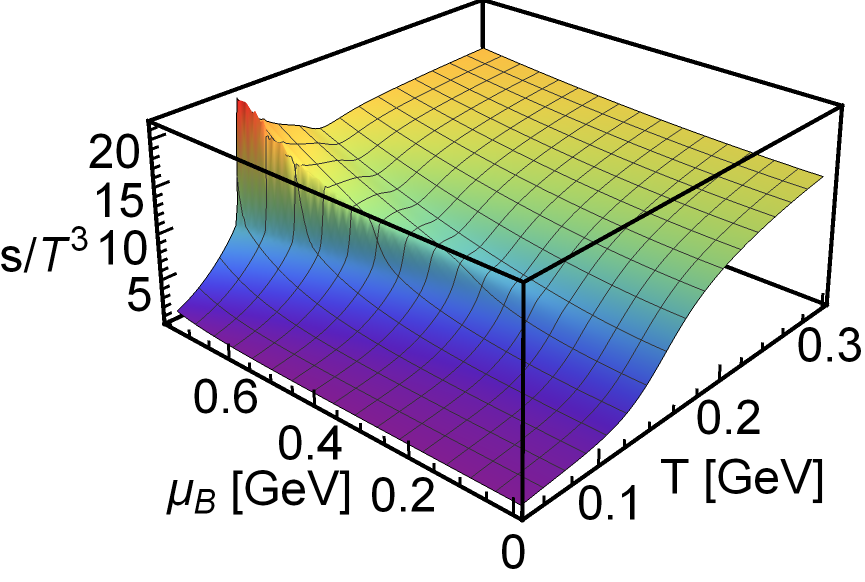} \hspace*{1mm}
  \includegraphics[width=0.22\textwidth]{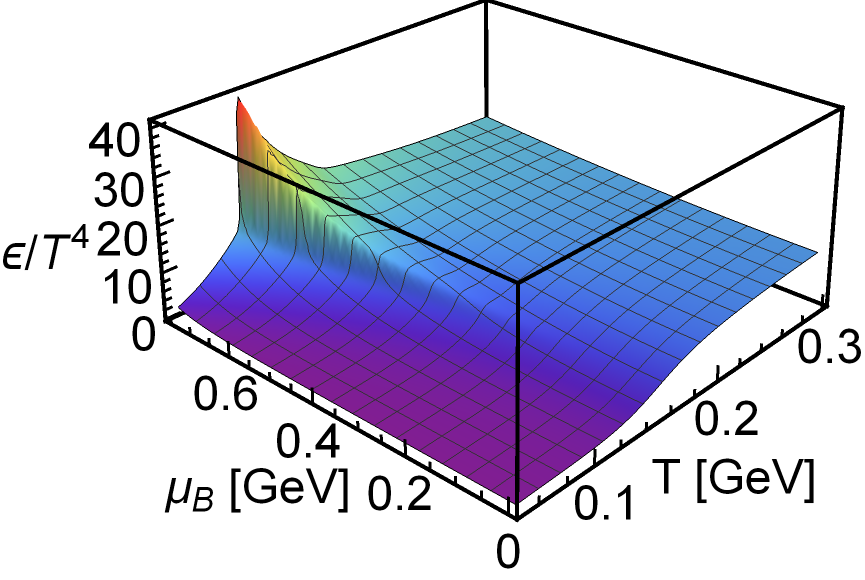}\\
\vspace*{-2mm}
  \caption{Calculated 3D plots for the pressure, number density, entropy density and energy density in terms of the temperature and chemical potential, normalized to a dimensionless form by the temperature $T$ with the respective power.}\label{fig:3dplot}
\end{figure}

We first consider the 3 flavor degenerate case with $\mu_{u} =\mu_{d} = \mu_{s} =\frac{\mu_{B}^{}}{3}$ which is the conventional case often applied in the hydrodynamics simulations.  With the obtained pressure $P(T,\mu_{B}^{})$ and number density $n_{q}(T,\mu_{B}^{})$ and using thermodynamic relations, one can compute the entropy density $s=\partial P/\partial T$, the energy density $\varepsilon= Ts - P + \mu_{B}^{} n_{B}$ and so on. The speed of sound squared $c_{s}^{2}$ is then given by   $c_{s}^{2} =\frac{\partial  P}{\partial \varepsilon}$. We show thus the 3-dimensional (3D) plot in terms of temperature and chemical potential for the pressure, energy density, number density and speed of sound in Fig.~\ref{fig:3dplot}.
The current results are consistent with the phase transition line obtained by lattice QCD simulations and functional QCD methods with a CEP at $(T,\mu_{B}^{})=(118,600)$~MeV.

\begin{figure}[t]
  \centering
  \includegraphics[width=0.4\textwidth]{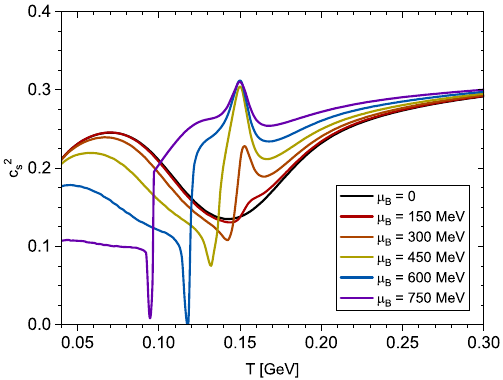}\\
\vspace*{-3mm}
\caption{Calculated speed of sound squared as a function of temperature $T$ at several values of the baryon chemical potential $\mu_{B}^{}$.}\label{fig:cs2Tmu}
\end{figure}

Now one may take a closer look at the EoS at different chemical potentials as depicted in Fig.~\ref{fig:cs2Tmu}.
For small chemical potentials, the speed of sound is a smooth function with a minimum around $c^{2}_{s}\sim$ 0.12 at phase transition point.
As the chemical potential becomes larger, the speed of sound becomes more oscillated near the transition temperature.
At the CEP and the first order phase transition region, the speed of sound at phase transition point becomes zero.
For the first order phase transition, since one has $\textrm{d}P=0$ and a finite $\textrm{d}\varepsilon$ during the phase transition,
the curve becomes discontinuous for the two phases.
The speed of sound in the vicinity of phase transition point is still finite, and only a few points reach to zero drastically.
More interestingly, as the chemical potential increases, there  appears a peak nearly above the phase transition temperature.
The maximum value of the peak gradually grows and saturates to the conformal limit $c_{s}^{2}=1/3$, which implies a  new feature of the QCD EoS  at high densities that may shed light on the studies of neutron stars~\cite{Xu:2009vi,Li:2022cfd,Li:2022okx}.

\begin{figure}[t]
\includegraphics[width=0.4\textwidth]{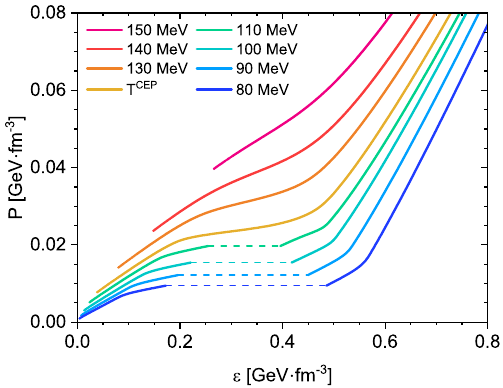}
\caption{Obtained pressure as a function of the energy density at several given temperatures; the Maxwell construction is displayed as the dashed lines.}\label{fig:pmaxwell}
\end{figure}

In the ideal case of the first order phase transition, there exists discontinuity in the number density, entropy density and also energy density. The discontinuity would be affected by the non-equilibrium effects in the dynamical evolution and will cause problems in the hydrodynamics simulation.
Here, we consider the Maxwell construction to fill the discontinuity with simply a linear transition of the thermodynamic functions from one phase to the other.
This construction leads to a vanishing speed of sound during the first order phase transition. To consider the spinodal decomposition, one may need a construction described in Ref.~\cite{Parotto:2018pwx}, and the speed of sound will be negative in the instable region.
In Fig.~\ref{fig:pmaxwell}, we show the construction for the pressure as the function of energy density. Using this construction, one can convert the thermodynamic quantities from the $(T,\mu_B)$ dependence into the $(\epsilon,n_B)$ dependence, which is then accessible for the hydrodynamic simulations.
It also  needs to mention that based on the current feature of EoS, a large value of $c^{2}_{s}$ near conformal limit does not necessarily mean that  there is no first order phase transition in neutron star as indicated in Ref.~\cite{Brandes:2023hma},
on the contrary, it might be an important feature when the QCD matter approaches the first order phase transition region at large chemical potential.
The EoS data as a function of  $T$ and $\mu_B$ together with the MUSIC input format for both B and BS cases are available in the github~(\href{https://github.com/YiLu1048576/fQCD-EoS-PhaseDiagramMap.git}{fQCD-EoS-PhaseDiagramMap}).

At last,  we check the conserved charge conditions satisfied in the heavy-ion collisions. Consider the 3 light flavors $f=u,d,s$, the conditions are:
\begin{equation}\label{eq:chrgCond}
  n_{S} = 0, \quad n_{Q}/n_{B} = r,
\end{equation}
with conserved charges $n_{B,Q,S}^{}$ which stand for the baryon, electric and strangeness densities, respectively, and $r$ the charge-to-mass ratio of the collided nuclei, e.g. $r_{\textrm{Au+Au}}^{} \approx 0.4$.

\begin{figure}[t]
  \centering  \includegraphics[width=0.44\textwidth]{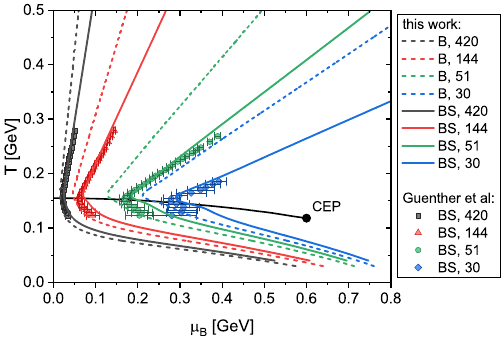}
  \caption{Calculated isentropic trajectories at $s/n_B$ = 420, 144, 51 and 30, with the 3-flavor-degenerate EoS~(B) and the charge-conserved EoS~(BS). The scatter points are the trajectories obtained from lattice calculation~\cite{Guenther:2017hnx} at these $s/n_{B}$ values.}\label{fig:isen}
\end{figure}

In this sense the 3-flavor-degenerate scenario is equivalent to $\mu_{Q}^{} =\mu_{S}^{} =0$ in Eqs.~\labelcref{eq:BS_muu,eq:BS_mud,eq:BS_mus}, i.e. only the baryon chemical potential is considered. On the other hand, for the (2+1)-flavor scenario Eq.~(\ref{eq:chrgCond}) is equivalent to:
\begin{gather}
  n_{s}(T,\mu_{s}) = 0, \label{eq:BS_0}  \\[2mm]
  \frac{2\,n_{u}(T,\mu_{u}^{}) - n_{d}(T,\mu_{d}^{})}{n_{u}(T,\mu_{u}^{}) + n_{d}(T,\mu_{d}^{})} = r. \label{eq:BS_1}
\end{gather}
Within our framework Eq.~(\ref{eq:BS_0}) requires $\mu_s = 0$. Therefore, for the charge conserved case the remaining task is to consistently solve Eqs.~(\ref{eq:BS_muu}), (\ref{eq:BS_mud}), (\ref{eq:BS_0}) and (\ref{eq:BS_1}). The parameters in Table~\ref{tab:param_m} is adapted for both $u$ and $d$ quark densities in Eq.~(\ref{eq:BS_1}).

we  show the calculated isentropic trajectories with $s/n_B = const.$ in Fig.~\ref{fig:isen}, for the 3-flavor-degenerate case which is denoted as B~(baryon), and the charge-conserved case with the constraint of the conserved charge conditions which is denoted as BS~(baryon-strange). The obtained isentropic trajectories are in good agreement of the previous studies. It shows that the strangeness neutrality pushes the trajectories into higher chemical potential region, which may have a impact on the initial conditions of heavy ion collisions~\cite{Fu:2018swz,Noronha-Hostler:2019ayj,Jiang:2021fun,Sun:2022cxp}.


\section{Application in hydrodynamic simulation}\label{sec4}

To evaluate the applicability of the constructed fQCD EoS in realistic dynamical evolution, we carry out the hydrodynamic simulation with the EoS as input.
In our hydrodynamic simulations, we implement the hydrodynamic model MUSIC~\cite{Schenke:2010nt, Schenke:2011bn} with the initial condition generated by AMPT~\cite{Lin:2004en}, where the initial energy/baryon number distribution are extracted at constant proper time $\tau_{0}$.
%
%
To compare the difference between various EoSs, we calculate the observables of thermal particles without hadronic scattering or decay effects.
See Ref.~\cite{Fu:2020oxj} for more details.
Here we do the calculation with ideal hydro and neglect the initial flow, which do not affect the final particle yields obviously.

\begin{figure}[!htb]
\vspace*{-3mm}
\includegraphics[width=0.44\textwidth, trim={1cm 0 1cm 1cm}, clip]{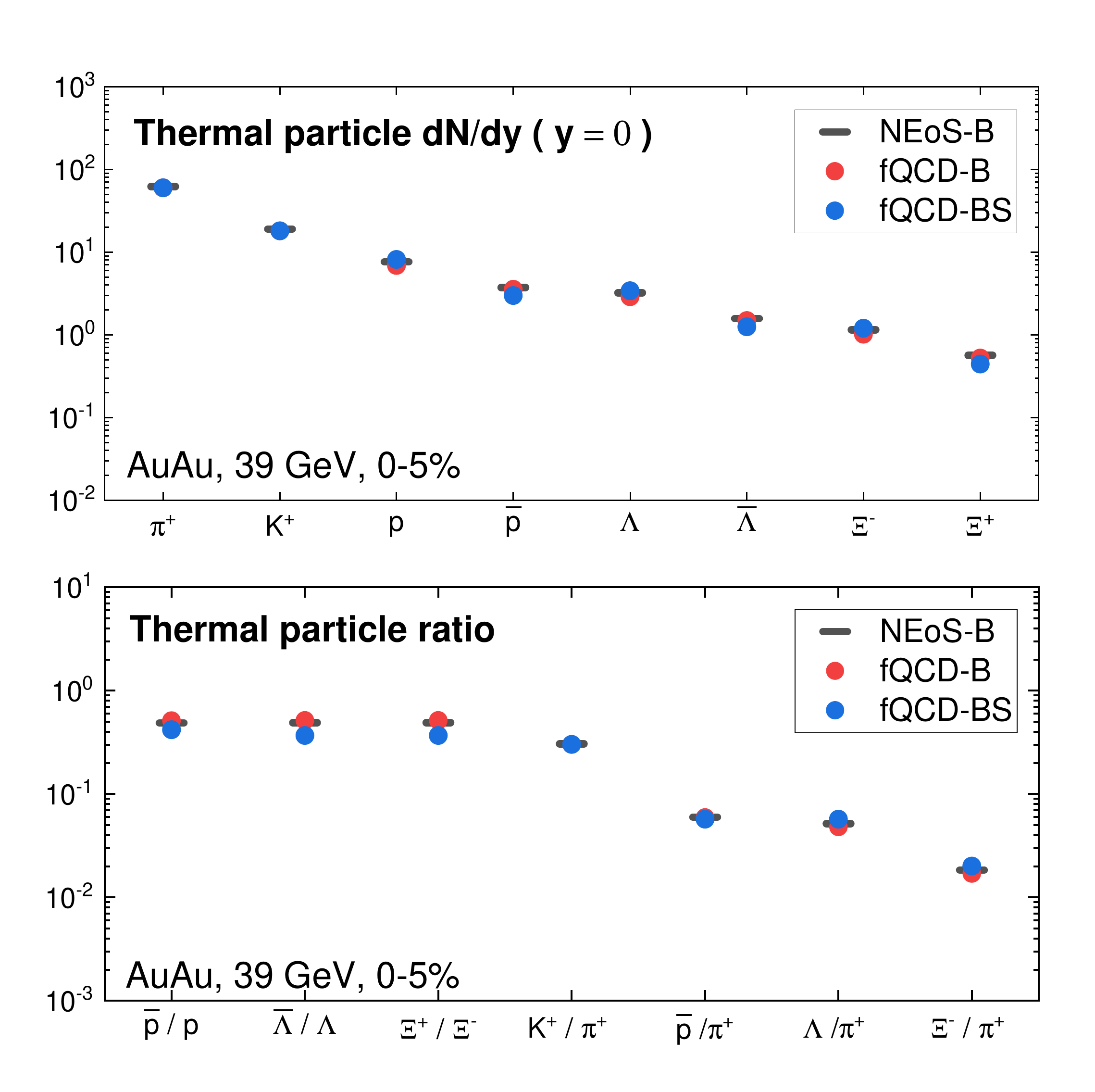}\\
\vspace*{-6mm}
\caption{\label{fig:yields} Identified thermal particle yields (upper panel) and their ratios (lower panel) in 0-5\% Au+Au collisions at $\sqrt{s_{NN}} = 39$~GeV. The scatters represent the model calculations with NEoS-B and fQCD equation of states on freeze-out surface, respectively.
}
\end{figure}

\begin{figure}[!tb]
\vspace*{-2mm}
\includegraphics[width=0.44\textwidth, trim={1cm 0 1cm 1cm}, clip]{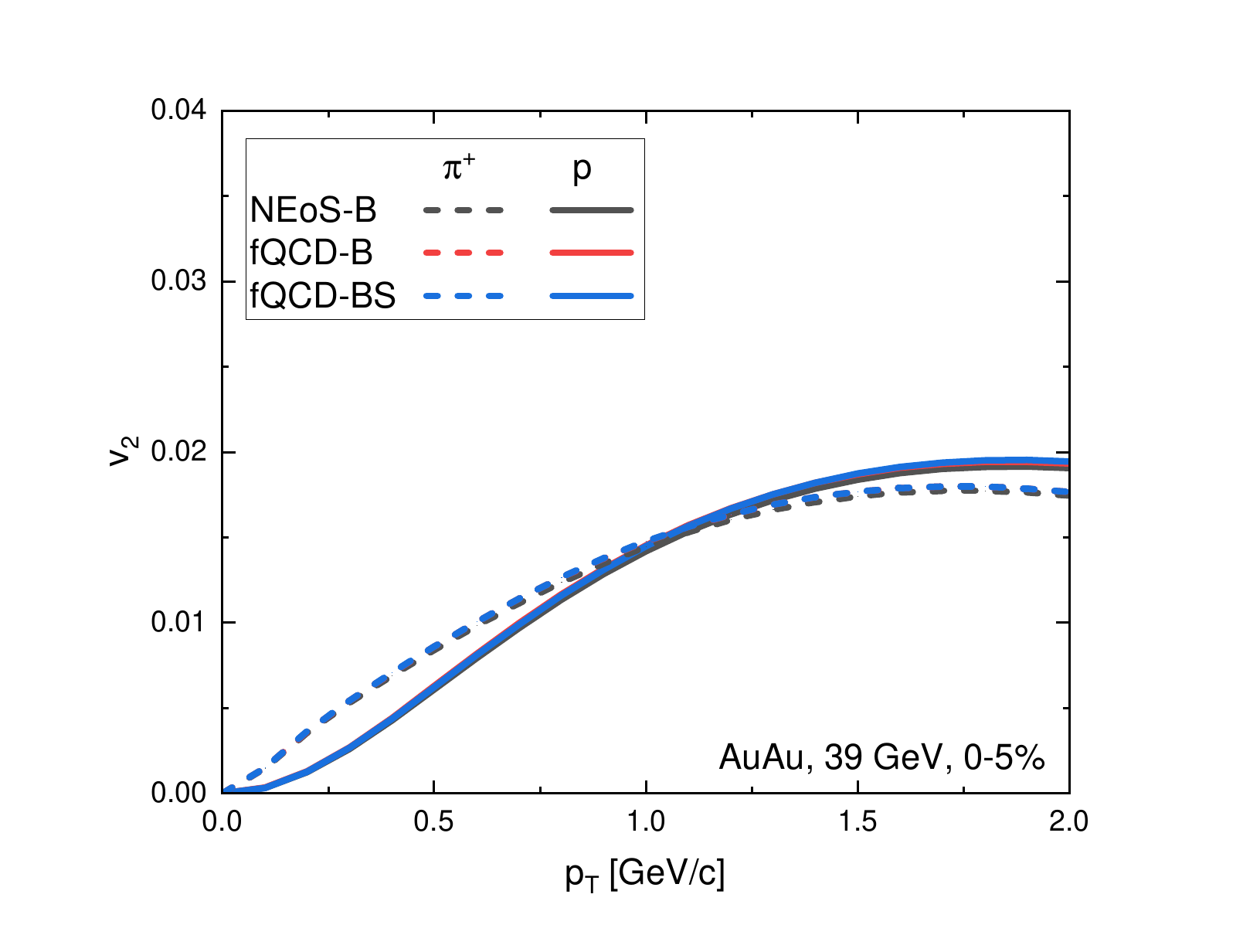}\\
\vspace*{-6mm}
\caption{\label{fig:v2} Differential elliptic flow $v_{2}^{}(p_{T})$ for $\pi^{+}$ and proton in 0-5\% Au+Au collisions at $\sqrt{s_{NN}} = 39$~GeV. The calculation is done on the hydrodynamic freeze-out surface. }
\end{figure}

Fig.~\ref{fig:yields} shows the identified particle yields and ratios calculated by AMPT + MUSIC in 0-5\% Au+Au collisions at $\sqrt{s_{NN}} = 39$~GeV.
As a comparison, we also show the results calculated by a crossover equation of state NEOS-B, which is based on lattice QCD simulation and hadron resonance gas model~\cite{Monnai:2019hkn}.
Note in most central Au+Au collisions at 39 GeV, the corresponding chemical freeze-out $(T, \mu_B)$ is around (156, 160) MeV, where fQCD-B and NEOS
show no obvious difference.
When considering the charge conservation in fQCD-BS, the anti-baryon/baryon ratio decreases due to larger baryon chemical potential, which also corresponds to the isentropic trajectories in Fig.~\ref{fig:isen}. The ratio for hyperon  like $\Lambda$ and $\Xi$ also show similar behavior.  Note that here the fQCD EoS-BS changes the anti-baryon/baryon ration, which may be further improved by considering the flavor mixing effect in the future.
%
In addition, Fig.~\ref{fig:v2} shows the corresponding elliptic flow $v_{2}$ of $\pi^{+}$ and proton calculated by pure hydrodynamics.
It shows evidently that, at the crossover region, changing EoS will not make obvious difference to the bulk evolution, which is also consistent with the result of Fig.~\ref{fig:isen}.

\begin{figure}[!htb]
\vspace*{-6mm}
\includegraphics[width=0.44\textwidth, trim={1cm 0 1cm 0}, clip]{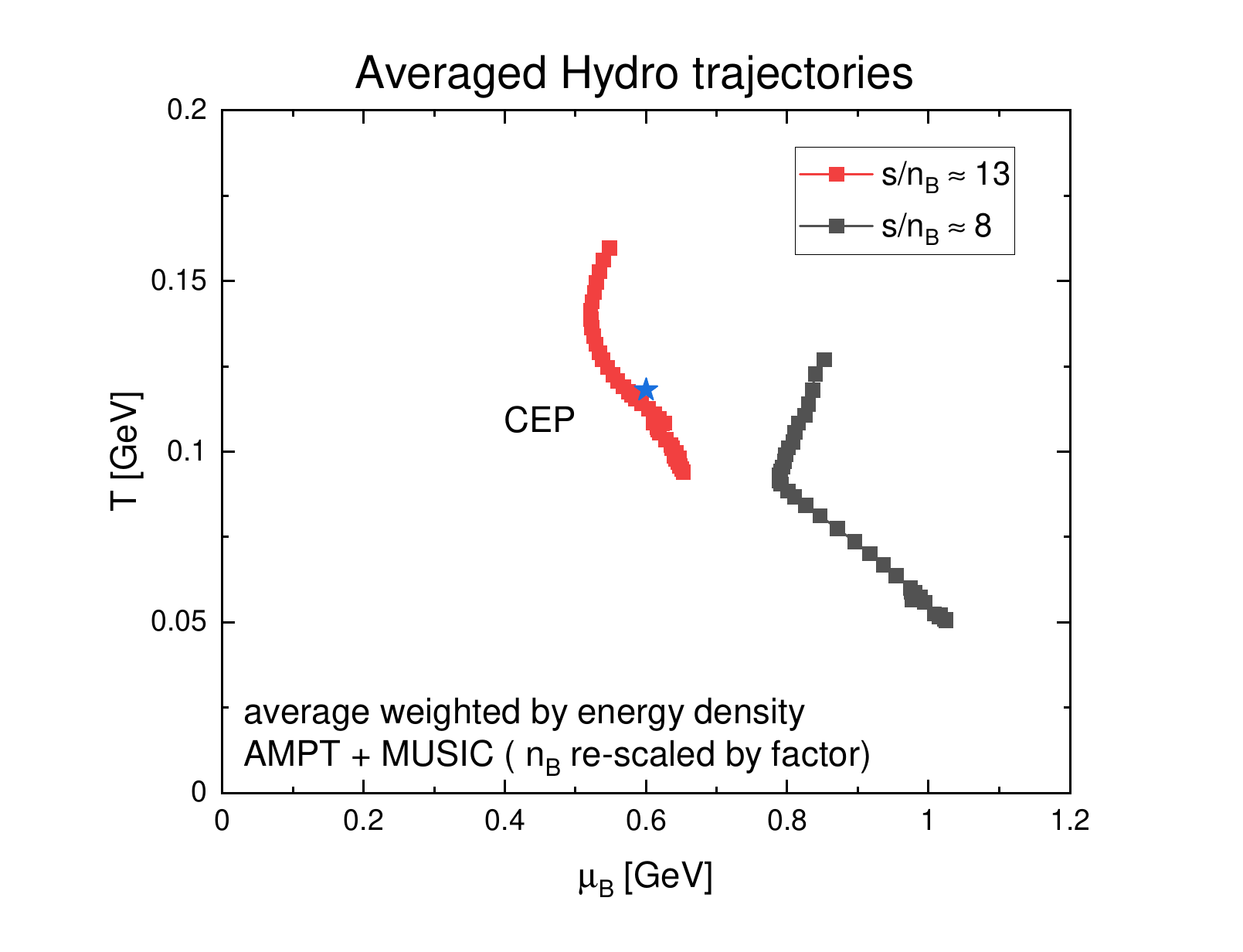}\\
\vspace*{-6mm}
\caption{\label{fig:hydro} Hydrodynamic trajectories at high baryon number density region. The scatters shows the hydro evolution with different initial $s/n$, where they are averaged in each time step with energy density weight. The blue ``star" shows the critical end point locate at $(T, \mu_B) = (118, 600)$~MeV. }
\end{figure}

To study the stability of the fQCD EoS near the first order phase transition, we run the hydrodynamics at higher baryon chemical potential region without freezeout.
Fig.~\ref{fig:hydro} shows the obtained averaged hydrodynamic trajectories at such baryon-rich region, where we take the AMPT initial profile at $\sqrt{s_{NN}} = 7.7$~GeV but re-scaled the initial net baryon density distribution by factors to match $s/n_B \approx 13$ and $s/n_B \approx 8$ in initial conditions, respectively.
In both cases, hydrodynamics start with same initial energy profile and can run stably for more than 10 fm/c.
Note for the case $s/n_{B} \approx 8$, the theoretical isentropic line will experience a first order phase transition with decreasing temperature. While in hydrodynamic simulation, the speed of sound will turn to zero at phase transition, so it cannot evolve across the phase transition line but go along with it instead.
How the system evolves after first phase transition is out of the reach of current hydrodynamic model, a more realistic dynamical description applying the constructed EoS with CEP and first order phase transition will be investigated in the future.

\section{summary}\label{sec5}

We proposed an improved construction on the QCD EoS in a functional QCD based scheme, which takes input from the lattice EoS at zero density, and moreover  utilizes the knowledge of the QCD phase diagram at finite density from functional QCD approaches, i.e. the phase transition line and the location of the CEP. By implementing the zero-momentum approximation,  the quark number density is expressed analytically, which ensures then that the EoS can be analytically calculated and is convenient for further applications.

We computed the thermodynamic quantities such as the pressure, the energy density and the entropy density, and eventually the speed of sound as a function of temperature and chemical potential. For small chemical potentials, the obtained minimum around $c^{2}_{s}\sim 0.12$ at the phase transition point is consistent with the results from lattice QCD simulation. As the chemical potential increases, the speed of sound drops drastically to zero at the phase transition point.  Moreover, the speed of sound at large chemical potential shows a conformal limit behavior right after the first order phase transition, which may have some impacts on neutron star properties.

 We applied our constructed EoS to get the isentropic trajectories. Our results are in a good agreement with those from lattice QCD simulations.  Besides, the current estimation gives that the evolution goes into first order phase transition region at   $s/n_B\sim 10$. This provides a benchmark for the QCD chiral phase transition in heavy ion collisions and the possible  new  phenomena such as the non-monotonicity of the triton yield ratio. The further investigation  either resolve the conflicts between theory and experiment, or reveal some new features of QCD.

 We also put the EoS into the hydrodynamic simulations.  For Au+Au collisions at $\sqrt{s_{NN}}=39$ GeV,  our results on particle yields, their ratios and also the elliptic flow of pion and proton are in good agreement with the results from a commonly used EoS.  Moreover,  we showed that with the current EoS,  it is possible  for the hydrodynamic simulations  to  reach the first order phase transition region.
A more realistic dynamical description of the first order phase transition will be investigated in the future.

\section{Acknowledgements}
 FG and YL thank the other members of the  fQCD collaboration~\cite{fQCD}  for discussions and collaboration on related subjects. This work  is supported by the National Natural Science Foundation of China under Grants  No. 12247107, No. 12175007 and No.~12075007.  BCF is also supported by the National Natural Science Foundation of China under Grants No. 12147173.   FG is supported by the National  Science Foundation of China under Grants  No. 12305134.

\bibliography{eos_map}

\end{document}